**Directional single-photon emission from deterministic quantum dot waveguide structures**

*Paweł Mrowiński,[1,2,*] Peter Schnauber,[1] Arsenty Kaganskiy,[1] Johannes Schall,[1] Sven Burger,[3] Sven Rodt[1] and Stephan Reitzenstein[1]*

[1] Institut für Festkörperphysik, Technische Universität Berlin, Hardenbergstraße 36, 10623 Berlin, Germany
[2] Laboratory for Optical Spectroscopy of Nanostructures, Department of Experimental Physics, Faculty of Fundamental Problems of Technology, Wrocław University of Science and Technology, Wyb. Wyspiańskiego 27, 50-370 Wrocław, Poland
[3] Zuse Institute Berlin, Takustraße 7, 14195 Berlin, Germany

*E-mail: pawel.mrowinski@pwr.edu.pl

Abstract

Chiral light-matter interaction can lead to directional emission of two-level light emitters in waveguides. This interesting physics effect has raised considerable attention in recent years especially in terms of on-chip quantum systems. In this context, our work focuses on tailoring single semiconductor quantum dot – waveguide (QD-WG) systems to emit single photons with high directionality. We use low-temperature in-situ electron-beam lithography enabled by cathodoluminescence mapping to select suitable QDs and to integrate them deterministically into linear waveguide structures at specific chiral points determined by numerical calculations. We observe excitonic and biexcitonic emission from the fabricated QD-WG structure in a confocal µPL setup enabling the optical characterization in terms of directional emission of circularly polarized photons emitted by integrated QDs. Our results show a high degree of anisotropy on the level of 54% for directional QD emission and antibunching in autocorrelation experiment confirming the fabricated QD-WG system, which is a prerequisite for using this effect in advanced applications in integrated quantum circuits. .

## 1. Introduction

Photonic waveguides with integrated quantum dots (QDs) are promising building blocks for the realization of quantum-photonic circuits to implement for instance Boson sampling and quantum gates for entanglement purification [1–8]. Interestingly, for propagating modes of light in waveguides, where light is transversally confined, a unique property of spin-momentum locking can be found [9,10], which links the local polarization (transverse spin angular momentum) and the propagation direction. This interaction is the basis for chiral light-matter coupling effects with quantum emitters with many exciting opportunities to control the direction of single-photon emission and absorption in highly functional integrated quantum technology [11,12]. It is worth mentioning that these opportunities are not only restricted to direct bandgap semiconductors realized in the GaAs/AlGaAs platform, but are also compatible with quantum emitters in heterogeneous GaAs/$Si_3N_4$ structures to couple photons into complex low-loss Si-based chips [6,13]. In this context, two-level systems providing well-defined polarized optical transitions, such as excitonic complexes in a QD, are most suitable for the integration into on-chip waveguide systems. By integrating such quantum emitters with waveguide systems one can achieve non-reciprocal, unidirectional photon emission and absorption, to implement for instance optical circulators for single photons, single-photon controlled switches, CNOT gates or complete spin networks [10].

In our previous work we developed the necessary technology platform to systematically study directional emission from a single InGaAs QD embedded deterministically into ridge waveguide structures using charged excitonic states to confirm chiral coupling existing in multimode WGs [14]. In this article we provided both experimental results for highly off-center positioned QDs consistent with the numerical results for few quasi-TE modes propagating along the WG. In the present work, by using another sample based on the same wafer material, we prove the high repeatability of our technology on the one hand, and on the other hand we go beyond existing results by demonstrating the quantum nature of highly directional emission in our QD-WG structures. The applied in-situ electron beam lithography (EBL) fabrication process allows for a pre-characterization of the sample to select spectrally and spatially isolated, optically bright QDs, to subsequently integrate them into linear waveguides terminated e.g. with grating outcouplers [5]. Using this in-situ EBL process we can place single QDs at well-defined off-center positions which, according to numerical simulations, exhibit maximum chiral light-matter interaction. The optical properties of deterministically processed QD-WG systems are studied via a spatially resolved (confocal) two-beam micro-photoluminescence (µPL) spectroscopy. In particular, we examine and compare the waveguide-coupled excitonic emission of deterministically integrated QDs via grating outcouplers from both ends of the linear waveguide. From the experimental data we evaluate the polarized emission contrast related to directional emission induced by the chiral light-matter coupling. Moreover, the quantum nature of emission and directional coupling is studied by photon autocorrelation measurements from the outcoupler output of such a QD-waveguide system.

## 2. Numerical simulations

Directional emission in the studied ridge waveguides is expected for in-plane circularly polarized dipole emission, i.e. helicity of $\pm 1$, in highly off-center position of the integrated QDs. In order to determine the exact lateral displacement $\Delta x$, we perform numerical simulations based on Maxwell's equations using a higher-order finite element method (FEM) implemented in the JCMsuite solver [15]. A 3D model of the dipole emission is schematically presented in Fig. 1a), together with the waveguide cross-section in Fig. 1b) showing an 800 nm wide and 800 nm high WG structure, where dark (bright) regions corresponds to GaAs (AlGaAs) layer forming distributed Bragg reflector (DBR) acting effectively as a cladding layer allowing for efficient wave guiding. After solving the problem for the preselected QD emission wavelength of 910 nm, we analyze the dipole coupling to the TE-like WG modes (TM-like modes are not considered as out-of-plane polarized component does not contribute to the chiral coupling),

WILEY-VCH

which are calculated separately in a cross-sectional 2D model (see Fig. 1b)). The directional dipole coupling factor $\beta\pm$ is evaluated by integrating the overlap between the dipole field distributions on both sides of the waveguide (see Fig. 1a)) with TE-modes by adding up all contributions. In the next step, we calculate the degree of anisotropy between left (-) and right (+) propagation direction in the waveguide, which is given by $(\beta_{\pm} - \beta_{\mp})/(\beta_{\pm} + \beta_{\mp})$ for a given right-hand or left-hand circular polarization of the dipole. Noteworthy, in the following experimental part the calculated degree of anisotropy is compared to the degree of circular polarization $DCP = (I_{\pm} - I_{\mp})/(I_{\pm} + I_{\mp})$, where $I_{\pm}$ is the intensity measured for right-hand (+) and left-hand (-) circular polarization for a single waveguide output, respectively. The latter is equivalent to the alternative approach using the selected polarization ($I_+$ or $I_-$) on both waveguide outputs, but this one can suffer from irregularities along the waveguide in experiment, which is not the case in the first case. In the Fig. 1c) the calculated degree of anisotropy is shown in dependence on the dipole position. This anisotropy shows characteristic oscillations due to higher order mode contributions below the displacement of Δx = 300 nm, and the maximum value of ~ 0.9 occurs at the chiral point with a displacement of about 340 nm [14].

### 3. Deterministic sample fabrication

The QD-waveguide system is based on a semiconductor heterostructure grown by metal-organic chemical vapor deposition (MOCVD) on (100) GaAs substrate. The epitaxial growth starts with a 300 nm thick GaAs buffer layer and it is followed by 23 mirror pairs of AlGaAs/GaAs forming a DBR mirror whose 80 nm wide stopband is centered at 930 nm. In addition, the DBR mirror increases the intensity of the light emitted towards the top direction which is beneficial for the pre-selection of suitable QDs on the planar sample in the in-situ EBL process. Moreover, in case of the waveguide with single D-shaped outcouplers it enhances the photon extraction efficiency towards the collecting optics. The lower DBR is followed by the 230 nm thick GaAs one-$\lambda$ cavity containing a single layer of InGaAs QDs of ~$10^9$ cm$^{-2}$ areal density which is in the center of GaAs.

A set of ridge QD-waveguides are defined by using low-temperature in-situ EBL realized by a customized scanning electron microscope (SEM), which is extended with a Helium flow cryostat, a cathodoluminescence (CL) spectroscopy unit, and a homemade EBL pattern generator. The in-situ EBL fabrication process flow is presented schematically in Figs. 2a) and b). The process begins with spin-coating the sample with a 100 nm thick layer of dual tone EBL resist CSAR 62 (AR-P 6200). Then the sample is mounted to the cold finger of the SEM cryostat and is cooled down to 10 K [16]. At this temperature we perform 2D CL scanning of the sample's surface to select suitable QDs based on their CL intensity, emission wavelength (target wavelength about 910 nm), spatial position as well as spectral pattern. This CL mapping process is performed using a homogenous electron dose of 10 mC/cm$^2$ (50 ms exposure time, 0.5 nA beam-current) with a grid size of 500 nm. As a result, the resist behavior changes to soluble in the 10 μm x 20 μm mapping area. In the subsequent low-temperature EBL step, a linear WG and grating outcouplers are patterned with high alignment accuracy of 30-40 nm [17] with respect to a pre-selected QDs in each mapping field. In this step we apply grey-scale lithography with a homemade proximity correction procedure [14] to avoid irregularities at the waveguide edges. For the EBL step the electron dose is increased to 50 mC/cm$^2$, so that the exposed areas become insoluble again [18]. These areas are maintained in the subsequent development process (at 300 K in the cleanroom) and act as etch masks in the final plasma enhanced reactive ion etching step in which the deterministically patterned QD-WG structure is transferred into the semiconductor material. An SEM image of a QD-WG structure fabricated deterministically by in-situ EBL is presented in Fig. 2c).

### 4. Experimental results

First, we present spectroscopic data obtained during the in-situ EBL process and from subsequent process evaluation. Fig. 3a) presents a CL map which shows pronounced local luminescence from a single QD at a wavelength of $(908.0 \pm 0.5)$ nm. The associated CL spectrum from this QD (QD position: x = $(9.13 \pm 0.03)$ μm, y = $(5.93 \pm 0.04)$ μm within a 9.5 μm x 9.5 μm mapping field) is depicted in Fig. 3b) – lower panel. Next, the CL spectrum is compared to a μPL spectra taken at 8-10 K from the corresponding QD-WG structure after processing for high (> 5 μW, red trace) and low (< 0.5 μW, black trace) excitation powers of 787 nm diode laser, as shown in Fig. 3b) – upper panel. This comparison confirms that the deterministic sample processing does neither influence the wavelength nor the spectral pattern of the selected QD in a significant way. This is an important point, which will become crucial for future developments targeting at more complex QD-WG circuits with multiple spectrally matched QDs. In the presented μPL studies, we observe saturation of the emission line at 908.13 nm suggesting recombination from neutral or charged excitonic state. The other lines are most probably originating from higher order excitonic states of the same QD, like biexciton, or from other QDs.

Next, we explore chiral light-matter coupling and the related directional emission of a QD-WG structure, which is expected at the chiral point [2,4] for highly off-center QD position, as indicated previously by numerical simulations. For this QD-WG with a width of $(825 \pm 25)$ nm the lateral displacement $\Delta x$ is set to $(350 \pm 25)$ nm during the in-situ EBL process. We investigate the DCP of top emission from the outcouplers. For this specific position for the normalized Stokes parameter one obtains $V/I = Im\,(E_x^* E_z^* + ExEz)/(E_x^2 + E_z^2) = \pm 1$ , where $E_x$ ($E_z$) is the time-dependent electric field in-plane vector component perpendicular (parallel) to the waveguide axis '$z$'. Accordingly, at the chiral point $DCP = \pm 1$ is naturally expected.

In Fig. 4 we present polarization-resolved μPL spectra for QD emission at 908.1 nm of the selected QD-WG structure. This QD transition is most probably a charged exciton state due to relatively narrow linewidth of ~70 μeV, as compared to the width of 140 μeV for excitonic states with a fine structure splitting of 70 μeV [14]. As a reference measurement, we first excite and detect polarization resolved μPL emission directly from the QD, which is located as schematically presented in the inset of Fig. 4. As expected, in this case we observe rather small DCP of (-0.06 ± 0.01). Next, by a raster-scanning we move the collection spot along the waveguide to the position of the left and right outcouplers, as shown in the inset of Fig. 4. In this case, we collect only outcoupled emission from the same QD excitonic states, which first couple to quasi-TE modes and propagate a distance of about 40 μm along the waveguide. In this case, we observe a pronounced DCP of (0.54 ± 0.01) for the right side and (-0.22 ± 0.01) for the left side of the waveguide. The opposite sign of DCP clearly confirms that in the deterministically fabricated QD-WG system with highly off-center QD position the chiral coupling, i.e. highly directional emission of circularly polarized photons through the waveguide, can be achieved. Nevertheless, in our experiment the DCP is limited to 0.54 and the absolute value at both sides is not equal. To our understanding this could be related to: (a) not ideal spatial matching of the QD position and the related transition dipole moment with the chiral position; (b) irregularities along the waveguide influencing mode structure; (c) outcoupler design is not fully polarization maintaining in principle; and (d) possible back-reflection from the opposite side of the waveguide. The last point has, however, rather minor effect in case of 80 μm long WG structure, as we expect the attenuation of directional emission of about 0.36 dB/μm and back-reflection of about 20%, which results in ~$10^3$ less intensity on the opposite side of the WG. Concerning our D-shape outcoupler design, it is expected that the initial in-plane polarization of the dipole in the waveguide, that couples mainly to TE-like guided modes, can change due to multimodal field interference which is a function of the emitter-outcoupler distance and also depends on the outcoupler's geometry. We verified that for the circularly polarized dipole at highly off-center position characterized by ~90% of directionality, which is close to chiral coupling for QD charged excitonic state, the outcoupled light has circularly polarized component of 48%, linearly polarized component of 15%, tilted (± 45 deg.) linearly polarized component of 18%, and unpolarized light component of 19%. This evaluation was made with a limited numerical aperture of 0.4 in the far-field of the top emission. In case of additional filtering applied in the experimental setup adjusted for a certain circular polarization state, which is realized by using a quarter-wave plate and a linear polarizer, the relevant circular polarization component is then highly sustained of ~74% (other components are reduced by a factor of 2), and therefore noticeably reducing the experimental error of directionality contrast evaluation.

For future applications in quantum-photonic circuits it is crucial to verify the quantum nature of directional emission. For this purpose, we measured the photon autocorrelation function g$^{(2)}$(τ) for the QD-WG structure under study. The experimental setup includes a fiber-coupled Hanbury-Brown and Twiss (HBT) configuration with two single-photon counting modules based on Silicon avalanche photodiodes. In Fig. 4a) we present the normalized g$^{(2)}$(τ) data taken for the emission directly from the QD position under continuous-wave non-resonant excitation 787 nm. We fit the raw data histogram with a model function $g^{(2)}(\tau) = 1 - (1 - g^{(2)}(0))exp(-\frac{|t|}{\tau})$ convoluted with the instrument response function (IRF) describing the timing resolution of the detection system given by $exp(-|t|/\tau_{IRF})$ and $\tau_{IRF} = 350\ ps$. In this way, we obtain $g^{(2)}(0)_{fit} = 0.25$ and $\tau = (2.2 \pm 0.2)\ ns$ for the convoluted function. By deconvolution with the IRF of the HBT setup, the corrected $g^{(2)}(0)_{dcv} = 0.20 \pm 0.06$, which is well below the limit of the single photon emission. Next, we measured the autocorrelation for the directional emission of the same QD exciton line detected at the right outcoupler. In this case, we obtain $g^{(2)}(0)_{fit} = 0.36$ and $\tau = (2.0 \pm 0.2)\ ns$ for the convoluted fit to the raw data, as it is presented in Fig. 4b), and deconvolution yields $g^{(2)}(0)_{dcv} = 0.25 \pm 0.06$. Noticeable antibunching given by $g^{(2)}(0)_{fit} < 0.5$ confirms that our deterministic chirally coupled QD-WG structure features directional coupling and emission at the single-photon level which is needed for further development of on-chip quantum optics and nanophotonics. In future, further reduction of the $g^{(2)}(0)$ value may be possible by using quasi-resonant p-shell excitation or strict resonant s-shell excitation of the QD exciton, which reduces uncorrelated background emission. Interestingly, in case of strict resonant excitation a high suppression of the laser scattering should be feasible [19] because of on-chip spatial filtering of the outcoupled emission collected with the confocal detection configuration at the grating outcoupler at the end of the linear waveguide.

## 5. Conclusions

Hereby, we presented the deterministic fabrication and optical study of a quantum dot waveguide structure serving as a directional single-photon source based on chiral light-matter interaction between the off-center quantum dot (displacement $\Delta x = (350 \pm 25)$ nm) and propagating modes (chiral point expected for $\Delta x \cong 340$ nm) of the multi-mode DBR ridge waveguide. Our device was fabricated via low-temperature in-situ EBL allowing us to integrate the QD in the target off-center position of the chiral point within 30-40 nm alignment accuracy. Chiral coupling depending on the spin configuration of excitonic state and helicity of emitted photons is verified by polarization resolved µPL measurements via high degree of polarization anisotropy of outcoupled emission observed from one end of the linear waveguide. Performing photon auto-correlation measurements we confirmed that quantum nature of directional emission in terms of $g^{(2)}(0) = 0.31 \pm 0.06 < 0.5$ is maintained in our chirally coupled QD-WG system. Our results on the controlled integration of single quantum emitters shows a high potential to pave the way for upscaling photonic waveguides to more complex on-chip integrated quantum circuits with novel quantum functionality.

**Acknowledgements**
This research was supported by the Polish Ministry of Science and Higher education within Mobilność Plus – V edycja. The research leading to these results has received also funding from

WILEY-VCH

the German Research Foundation through CRC 787 'Semiconductor Nanophotonics: Materials, Models, Devices' and from the European Research Council under the European Union's Seventh Framework ERC Grant Agreement No 615613.

Received: ((will be filled in by the editorial staff))
Revised: ((will be filled in by the editorial staff))
Published online: ((will be filled in by the editorial staff))

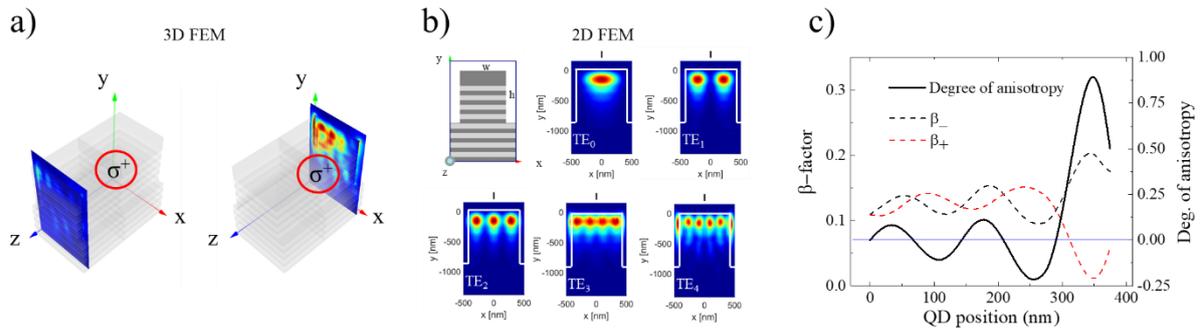

**Figure 1.** a) Calculated intensity profiles on both sides of the DBR-ridge-waveguide for circularly polarized dipole emission. High emission anisotropy is achieved due to the rather large lateral displacement $\Delta x = 350$ nm of the dipole (marked with red circle). b) Cross-sectional view of the DBR-waveguide structure together with the illustration of the first five quasi-TE propagating modes that significantly contribute to the chiral coupling with the off-center dipole emitter. c) Mode coupling $\beta$-factor representing directional emission with ($\pm$) notation related to both directions along the waveguide together with the degree of anisotropy in dependence on the QD (dipole) off-center position.

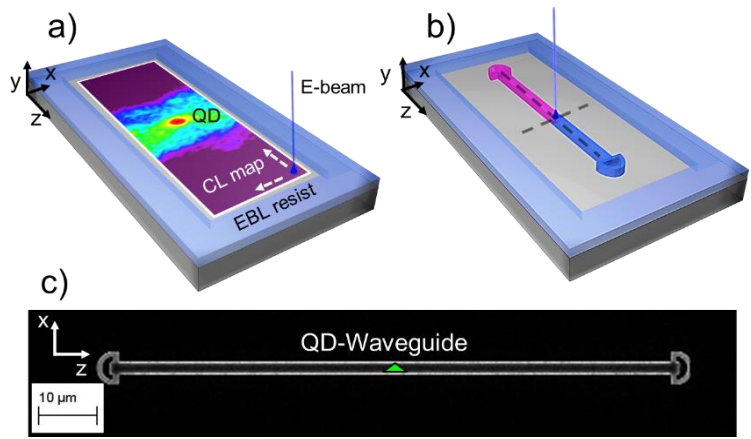

**Figure 2.** a) and b) Schematic view of in-situ electron-beam lithography processing. In the first step (a) a 2D cathodoluminescence map is collected over the writing field which allows one to select a bright single QD and to integrate it directly afterwards into a (b) the target waveguide structure with high alignment accuracy. c) SEM image of the fabricated QD-800 nm wide and 80 μm long waveguide with D-shaped outcouplers on both ends.

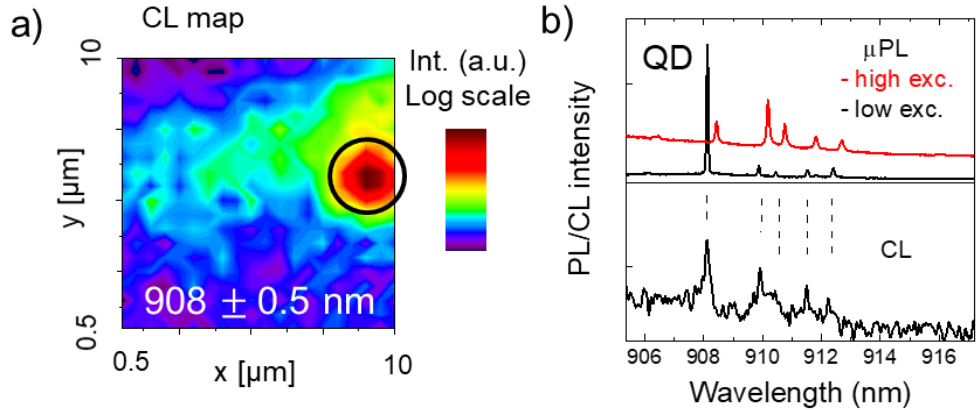

**Figure 3.** a) 2D cathodoluminescence map taken within a (908.0 ± 0.5) nm wavelength range to determine the position of a suitable QD. b) Comparison of CL and μPL spectral patterns from the selected QD before and after waveguide integration, respectively. The high-excitation power spectra in red color is red-shifted by 0.5 nm for clarity.

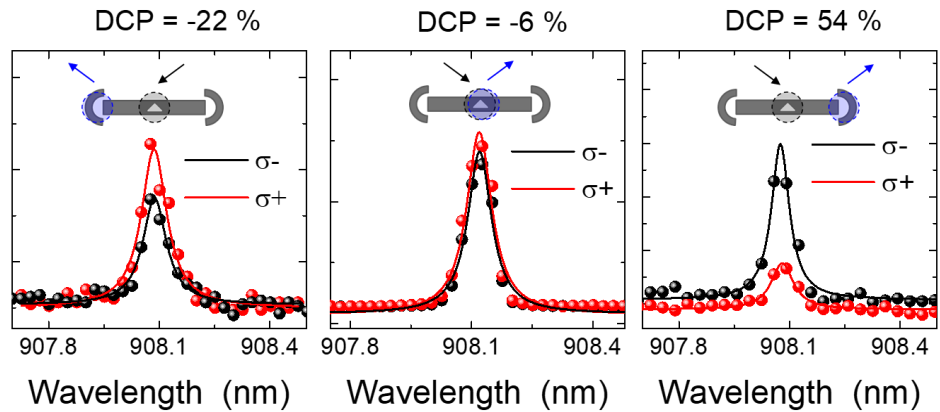

**Figure 4.** Polarization resolved μPL spectra of a QD-WG with Δx = (350 ± 25) nm displacement, and the degree of circular polarization (DCP) of emission determined directly at the QD position (center) and from both outcouplers (left/right). The opposite DCP sign on both sides of the WG confirms pronounced directionality of emission for the deterministically fabricated QD-WG structure.

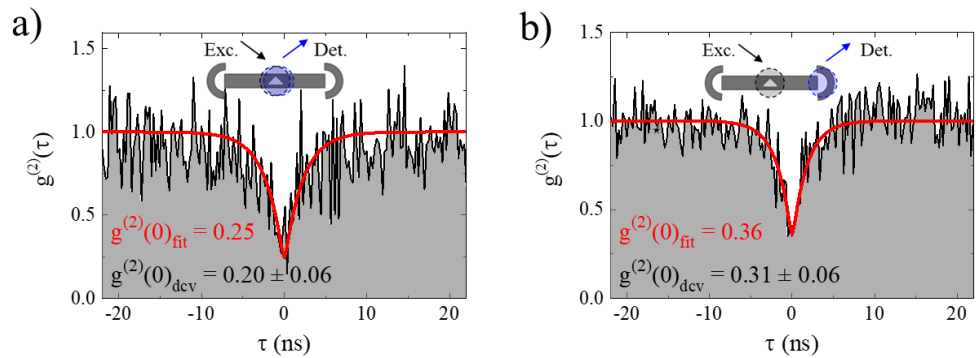

**Figure 5.** Photon autocorrelation measurement of the QD-WG with Δx = (350 ± 25) nm presented in Fig. 4. QD exciton emission coupled to waveguide modes shows antibunching for QD position in (a) and for outcoupled emission in (b) confirming the quantum nature of directional emission shown. The $g^{(2)}(0)_{fit}$ corresponds to the direct fit to the raw data including timing resolution of the setup, while $g^{(2)}(0)_{dcv}$ represents the deconvoluted value.